\documentclass[aps,pre,twocolumn,a4paper,superscriptaddress,amsmath,amssymb]{revtex4-2}

\usepackage{graphicx}
\usepackage{dcolumn}
\usepackage{bm}
\usepackage{textcomp}
\usepackage{hyperref}
\usepackage{siunitx}
\usepackage{mathtools}
\usepackage[thinc]{esdiff}
\usepackage{nicefrac}

\usepackage{hyperref}
\hypersetup{
    colorlinks=true,
    linkcolor=blue,    
    urlcolor=blue,
    citecolor=blue
    }

\usepackage{color}
\definecolor{g}{rgb}{0,1,0}
\definecolor{darkgreen}{rgb}{0,0.55,0}
\definecolor{darkred}{rgb}{0.4,0,0.1}
\definecolor{comment}{rgb}{0.0,0.25,0.45}

\begin{document}

\title{Collective dynamics of macroscopic photoactive matter under alternating excitation patterns}

\author{Sára \surname{Lévay}}
\email{slevay@unav.es}
\affiliation{Departamento de Física y Matemática Aplicada, Facultad de Ciencias, Universidad de Navarra, E-31080 Pamplona, Spain}
\author{Axel \surname{Katona}}
\affiliation{Departamento de Física y Matemática Aplicada, Facultad de Ciencias, Universidad de Navarra, E-31080 Pamplona, Spain}
\author{Raúl \surname{Cruz Hidalgo}}
\affiliation{Departamento de Física y Matemática Aplicada, Facultad de Ciencias, Universidad de Navarra, E-31080 Pamplona, Spain}
\author{Iker \surname{Zuriguel}}
\affiliation{Departamento de Física y Matemática Aplicada, Facultad de Ciencias, Universidad de Navarra, E-31080 Pamplona, Spain}

\begin{abstract}
We present experiments on the collective dynamics of macroscopic photoactive self-propelled particles subjected to spatiotemporally varying excitation. The particles move within an arena divided into two regions with different illumination intensities, creating alternating bright (more active) and dark (less active) zones. Under such conditions, the system exhibits a robust migration from the more active region toward the less active region, demonstrating a strong response to external modulation. This response depends sensitively on the frequency of the illumination pattern: at low frequencies, particles follow the changing landscape, whereas at higher frequencies, the response diminishes. We show that this behavior arises from the interplay between the imposed excitation and the intrinsic dynamics of the particle clusters that form spontaneously. To explain these features, we extend a kinetic model previously introduced in Lévay \textit{et al.} [\href{https://journals.aps.org/prl/abstract/10.1103/t7st-flj2}{Phys. Rev. Lett. \textbf{135}, 098301 (2025)}], hence revealing the most important parameters governing the transition between the responsive and unresponsive regimes.
\end{abstract} 

\date{\today}

\maketitle

\section{Introduction}

Active matter represents a class of out-of-equilibrium systems characterized by the collective behavior of numerous interacting self-propelled agents that transform energy into mechanical motion~\cite{baconnier2025self, tevrugt2025what}. Active agents can originate from natural sources, such as microbial organisms and animals, or can be artificially engineered, including colloidal particles and robotic systems. Also, they are diverse in scale (ranging from nanometer-sized entities to those measuring meters in length) and imply a wide range of interaction types, including physical contacts, hydrodynamic forces, magnetic forces, and social interactions. As a result, active matter has garnered substantial attention over recent decades, inspiring extensive research and applications across different systems including microrobots~\cite{palagi2016structured, hu2018small, xu2019millimeter, cui2019nanomagnetic, miskin2020electronically}, colloidal particles~\cite{wensink2008aggregation, buttinoni2013dynamical, palacci2013living, lozano2016phototaxis, ginot2018aggregation, vutukuri2020light}, bacterial systems~\cite{arlt2018painting, be2019statistical}, vibrated granular materials~\cite{martinez2009enhanced, scholz2018rotating, bar2020self, soni2020phases, mohammadi2020dynamics}, robotic swarms~\cite{oh2017bio}, animal groups~\cite{ariel2015locust, ginelli2015intermittent}, and pedestrian dynamics~\cite{corbetta2023physics}.

The motion of active particles arises either from intrinsic forces generated by the particles themselves or from external stimuli. The latter case, when the motion is a response to environmental cues, is called taxis. In particular, the terms phototaxis or photoactivity are used when the activity is driven by light, an instance case that has been extensively studied in systems composed of microscopic particles~\cite{palacci2013living, lozano2016phototaxis, palagi2016structured, arlt2018painting, miskin2020electronically, vutukuri2020light}. Here, we investigate photoactivity at the macroscopic scale, where centimeter-sized particles, which interact among them and the walls only through direct physical contact, are activated by light~\cite{siebers2023exploiting,levay2025cluster}. In this context, the system can be regarded as photoactive granular matter. The dynamics of the system can be manipulated by adjusting the intensity of illumination: enhanced light exposure increases the activity of the agents, while reduced illumination leads to diminished activity. 

The advantages of using this type of system are twofold. First, since agents are granular and their interactions occur exclusively through direct physical contact, hydrodynamic and social forces do not play a role in the dynamics. This characteristic provides us with a higher degree of control over experimental conditions, thus improving reproducibility, a challenge often encountered in microscopic systems. Second, the external modulation of the agents' activity allows testing the response of the system to both spatial and temporal gradients of activity. This flexibility facilitates the simulation of various scenarios, including gradient patterns and alternating illumination. The latter is, precisely, the objective of this research. 

Our macroscopic photoactive agents exhibit persistent directed motion with certain degree of chirality \cite{carrillo2025depinning} \nocite{SM} as the original hexbug nano\textregistered~\cite{hexbugs} in which they are based. These particles have been utilized in various studies addressing phenomena such as clogging~\cite{patterson2017clogging}, sorting~\cite{deblais2018boundaries, barois2020sorting, sinaasappel2025collecting}, traffic jams~\cite{barois2019characterization, patterson2023fundamental}, robotic superstructures~\cite{boudet2021collections}, resetting~\cite{altshuler2024environmental}, actuation in active solids~\cite{baconnier2022selective}, emergent behavior~\cite{xi2024emergent}, and inertial dynamics~\cite{leoni2020surfing}.

\begin{figure*}[t]
\centering
\includegraphics[width=0.95\textwidth]{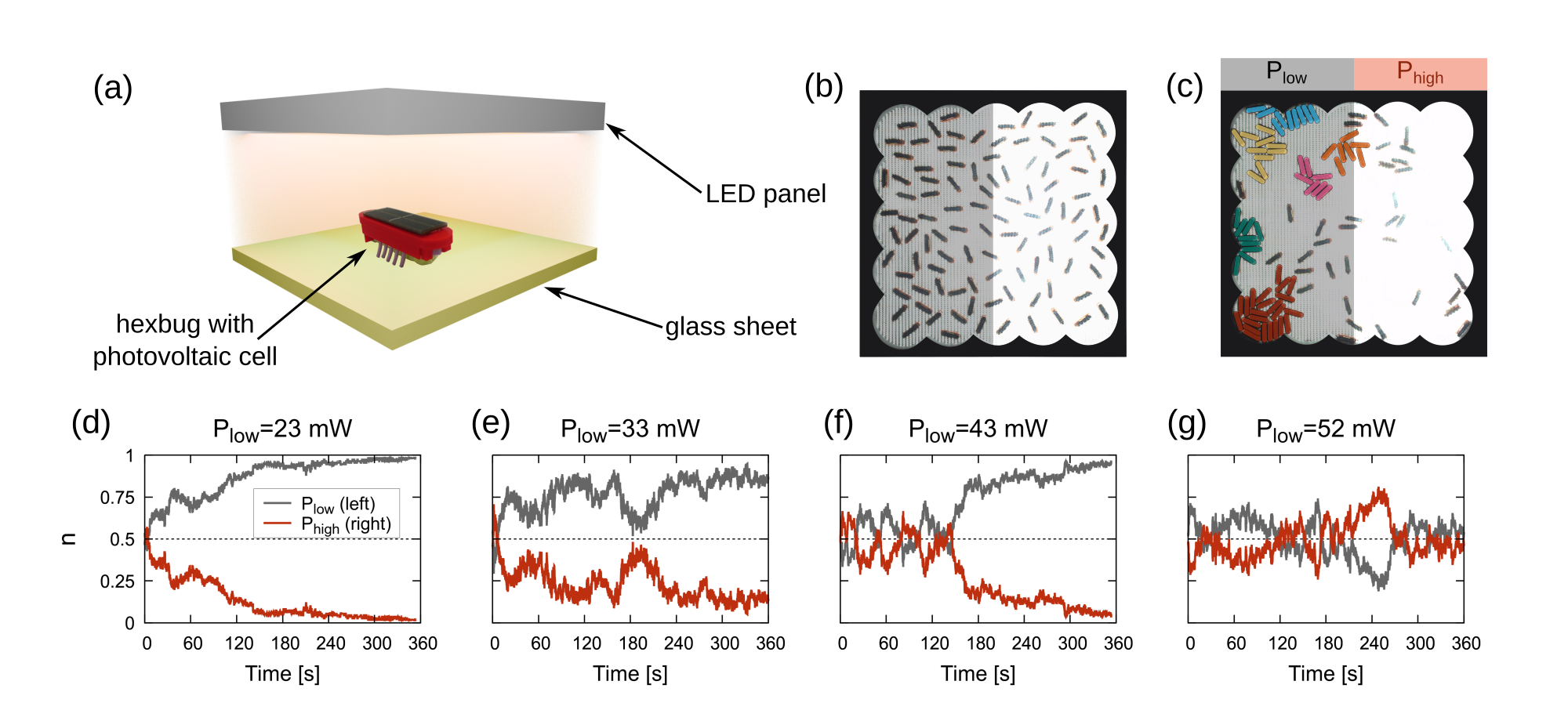}
\caption{\label{Fig:setup_sep}(a) Schematic representation of the experimental setup with the photoactive agent, a hexbug equipped with a photovoltaic cell, moving on a glass sheet, while the external control is attained by an LED panel. (b) Initial, random configuration of the agents. (c) Snapshot of the system: The right-hand side is illuminated with high intensity ($P_{\text{high}}=72$~mW, represented by red), while the left-hand side is illuminated with low intensity (represented by gray). Note that we have applied an additional gray shading on the left-hand side of the images to emphasize the different illumination of the two sides. Equally colored hexbugs belong to the same cluster, defined as groups of at least four agents that remain in contact during at least one second. (d)-(g) Time evolution of agent populations for both sides of the arena in experiments with stationary halved illumination. The right-hand side illumination is $P_{\text{high}}=72$~mW while the left-side illumination is always lower and different in each panel: (d) $P_{\text{low}}=23$~mW, (e) $P_{\text{low}}=33$~mW, (f) $P_{\text{low}}=43$~mW, and (g) $P_{\text{low}}=52$~mW. Supplemental Material (SM) videos 1-4~\cite{SM} represent these four example cases, respectively.}
\end{figure*}

Due to their shape and motion characteristics, particle-particle and particle-wall interactions lead to the formation of clusters. Recently, we have shown that under homogeneous illumination, thus homogeneous activity, a transition from unstable to stable clustering can be observed~\cite{levay2025cluster}. Specifically, we found that low particle populations combined with high activity lead to the formation of small clusters that tend to dissolve rapidly. On the contrary, high particle populations and low activity promote the emergence of stable clusters that endure throughout the whole duration of the experiment. Based on this knowledge, here we apply spatio-temporal alternation of high and low intensities, then investigating the system response depending on the frequency of this excitation.  

Our objective is to rigorously analyze the interaction among cluster formation, stability, and agent activity. This exploration is particularly pertinent in light of the increasing demand for precise control over the spatial and temporal distribution of active agents. Applications of this research include the strategic maintenance of specific agent densities in targeted locations, as well as the prevention of undesirable clustering, which can be crucial for effectively delivering loads to specific targets. Through this work, we seek to enhance the understanding of how variable activity landscapes influence agent dynamics and spatial organization.

The manuscript is organized as follows: In Sec.~\ref{Sec:setup}, we present the experimental setup and the applied protocols. Results obtained with a halved light intensity pattern stationary over time are presented in Sec.~\ref{Sec:stationary}, while the case of alternating activity patterns is discussed in Sec.~\ref{Sec:alternating}. In Sec.~\ref{Sec:model} we introduce an evolution of the previously presented kinetic model~\cite{levay2025cluster} describing the interplay of changing activity patterns and cluster formation. Finally, in the Conclusion and Perspectives section, we summarize our findings and discuss their relevance.


\section{Experimental setup and Methods}\label{Sec:setup}

The experimental setup is identical to the one presented in Ref.~\cite{levay2025cluster} with the uniqueness that here the illumination is not homogeneous in space and may also vary over time. Our photosensitive self-propelled particles [see Fig.~\ref{Fig:setup_sep}(a)] are an evolved version of the hexbug nano particles, which exhibit directed movement with certain degree of chirality~\cite{carrillo2025depinning}, although in this work we have selected only the less chiral particles. Each particle is equipped with ten asymmetric soft rubber legs and is activated by a vibrating motor, which, in our configuration, is powered by a photovoltaic cell. Under illumination, particles start moving with a velocity that depends on the vibration frequency~\cite{levay2025cluster}: the higher the illumination intensity, the higher the vibrating frequency and the larger the particle velocity. We stress that as hexbugs have no information about their surroundings, collisions are the only form of interaction, both between agents, and among the agents and the arena.

The particles move on a horizontal glass sheet with dimensions \numproduct{80x80}~cm$^2$ and are confined within a flower-shaped 3D printed plastic enclosure, referred to as the arena [see Fig.~\ref{Fig:setup_sep}(b) and \ref{Fig:setup_sep}(c)]. This design is specifically tailored to guide the particles toward the center and away from the boundaries~\cite{kumar2014flocking}. Before each experimental trial, the particles are randomly placed in varying positions and orientations within the arena, as illustrated in Fig.~\ref{Fig:setup_sep}(b). The illumination panel, also measuring \numproduct{80x80}~cm$^2$, comprises LED strips mounted on an aluminum plate positioned above the arena and connected to ESP32 microcontrollers, making the panel entirely programmable and enabling the application of illumination fields with temporal and spatial gradients. An Imaging Source DFK-37BUX250 camera (30 fps) is situated beneath the setup to capture video recordings of the particles’ movements. Subsequent analysis employs conventional image processing techniques combined with object detection algorithms to extract trajectory information for each agent.

Through this research, our objective is to gain insight into the collective dynamics of agents under varying illumination conditions. To this end, we implement two types of experiments, termed stationary halved illumination and alternating light intensity. In both cases, we place $N_T=100$ agents in the arena and apply two distinct intensity levels, $P_{\text{high}}$ and $P_{\text{low}}$, to opposite halves as illustrated in Fig.~\ref{Fig:setup_sep}(b) and \ref{Fig:setup_sep}(c). The intensity is quantified as the power received by the particles through the solar cell. $P=72$~mW represents the upper limit achievable with our setup, yielding particle velocities in the range of $12-14$~cm/s, while $P{=}23$~mW signifies the minimal illumination level that still permits particle mobility with a typical velocity of $8$~cm/s.

For the stationary halved experiments, the illumination pattern does not change over time. For these trials, we carry out several $6$-min-long experiments maintaining $P_{\text{high}}$ at $72$~mW and change $P_{\text{low}}$ over values of $23$, $33$, $43$, and $52$~mW. For each run, we track the hexbug trajectories and characterize the system dynamics which typically shows accumulation of particles in the lower-illuminated area. In the alternating case, we focus on a dynamic configuration where the illumination levels alternate between $P_{\text{high}}=72$~mW and $P_{\text{low}}=23$~mW at the two halves of the arena, with a frequency characterized by the switching period $T=$~\numlist{5;10;20;30;40;50;60;70;80}~s. A series of experimental runs, each lasting $10$ min, is executed for each frequency, then allowing the analysis of the system response.


\section{Stationary halved illumination}\label{Sec:stationary}

As explained above, we imposed a static excitation pattern in the arena, consisting of high illumination on the right side and low illumination on the left side [Fig.~\ref{Fig:setup_sep}(b) and \ref{Fig:setup_sep}(c)]. We then explore the system dynamics over the $6$ min experiments by analyzing the temporal evolution of particle populations on both sides. Figures~\ref{Fig:setup_sep}(d)–\ref{Fig:setup_sep}(g) show four representative runs for a fixed high-illumination level of $P_{\text{high}} = 72$~mW, combined with four distinct illumination levels on the darker side. The corresponding experimental movies are provided in the Supplemental Material (SM) videos 1–4~\cite{SM}.

\begin{figure}[t]
\centering
\includegraphics[width=0.5\textwidth]{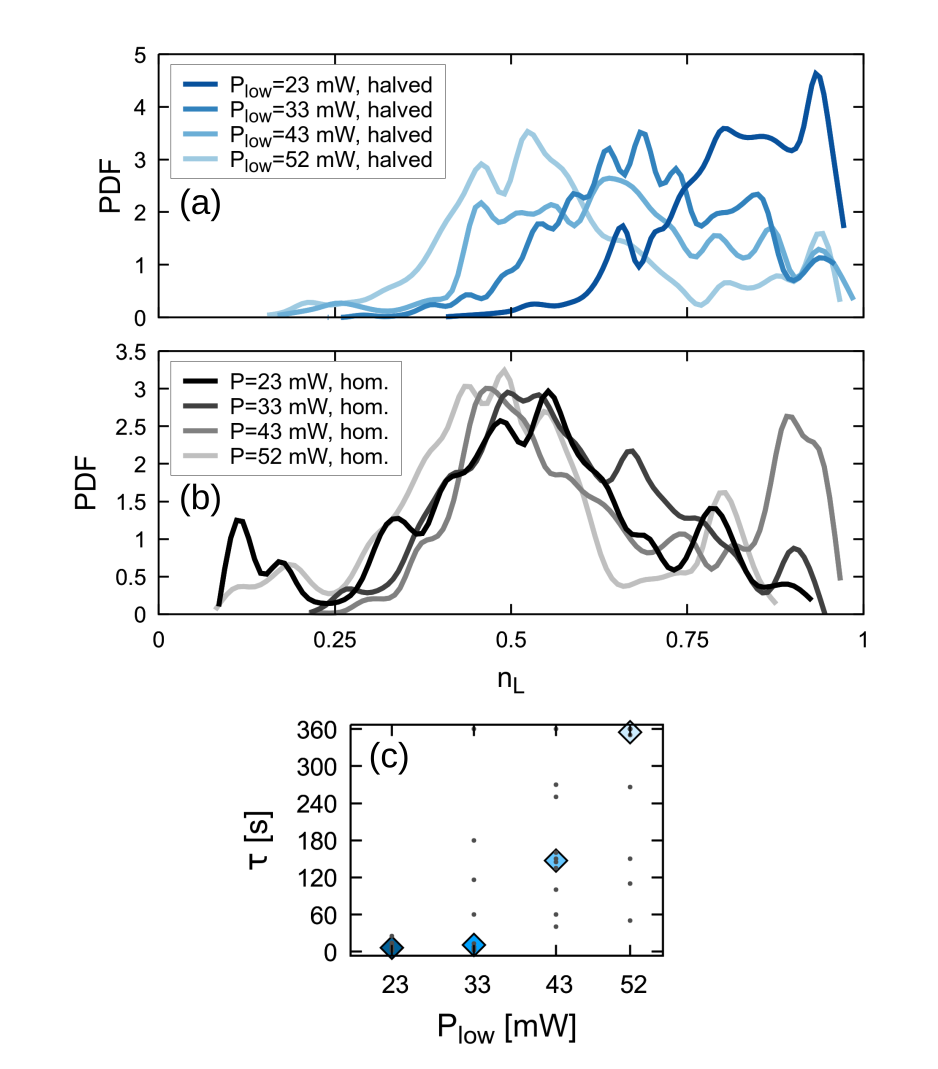}
\caption{\label{Fig:stationary}(a) Probability distribution functions (PDFs) of the relative population size obtained for the left (low-activity) side of the system ($n_L=N_L/N_T$) for levels of $P_{\text{low}}=23, 33, 43, 52$~mW for the stationary halved illumination. (The deeper the color, the lower the illumination.) The right (high-activity) side was always illuminated with $P_{\text{high}}=72$~mW. (b) PDFs of the relative population measured on the left-hand side when the arena is homogeneously illuminated in the experiments of Ref.~\cite{levay2025cluster}. To facilitate the comparison, we display the cases where the homogeneous illumination level is the same as $P_{\text{low}}$ of the halved case of panel (a). (c) Separation times measured in the stationary halved experiments corresponding to the four distinct $P_{\text{low}}$ levels. Each experiment was repeated ten times. Gray dots show separation times of individual experiments while blue diamonds represent the median.}
\end{figure}

At the lowest illumination levels on the dark side [Figs.~\ref{Fig:setup_sep}(d) and \ref{Fig:setup_sep}(e)], population separation occurs almost immediately. For $P_{\text{low}} = 23$~mW, nearly all particles migrate toward the darker side and remain there, with the relative population approaching one at long times. For $P_{\text{low}} = 33$~mW, the system appears to reach a stationary state in which the relative population fluctuates around $0.7$, occasionally exhibiting large fluctuations [e.g., the one near $t \approx 180$~s in Fig.~\ref{Fig:setup_sep}(e)]. A similar trend is observed for a slightly higher illumination level, $P_{\text{low}} = 43$~mW [Fig.~\ref{Fig:setup_sep}(f)], although in this case the initial separation is delayed, with populations remaining near $n = 0.5$ on both sides during the first $\sim 150$~s. For the case with the highest illumination at the dark side $P_{\text{low}} = 52$~mW [Fig.~\ref{Fig:setup_sep}(g)], no sustained population separation is observed: both sides remain close to $n = 0.5$ throughout the experiment. This outcome is likely due to the relatively small activity contrast between the two halves of the arena.

The behavior shown in Figs.~\ref{Fig:setup_sep}(d)-\ref{Fig:setup_sep}(g) corresponds to four representative examples of the dynamics taking place. A better exploration requires the analysis of the ten experiments repeated for each condition. To this end, in Fig.~\ref{Fig:stationary}(a) we represent the probability distribution function (PDF) of the fraction of particles located on the dark side of the arena for different values of $P_{\text{low}}$. For these plots we consider the whole temporal series, including the transient period. The differences between illumination levels are pronounced and consistent with the trends observed in Figs.~\ref{Fig:setup_sep}(d)–\ref{Fig:setup_sep}(g). For the largest $P_{\text{low}}$ (lightest blue), the distribution is only weakly asymmetric, exhibiting a slight bias toward higher relative populations on the darker side. As $P_{\text{low}}$ decreases (darker blues), the asymmetry increases. In the lowest-intensity case ($P_{\text{low}} = 23$~mW), the PDF displays a sharp peak near $n_{L} = 1$, indicating that nearly all particles accumulate on the dark (left) side.

For comparison, Fig.~\ref{Fig:stationary}(b) shows the PDFs obtained from experiments reported in Ref.~\cite{levay2025cluster}, where the arena was homogeneously illuminated. In this case, the PDFs are obtained counting the number of particles accumulating at the left-hand side for the same illumination levels used as $P_{\text{low}}$ in Fig.~\ref{Fig:stationary}(a). In all cases, the distributions exhibit a pronounced peak at $n_L = 0.5$, indicating equal likelihood of finding particles on either side of the uniformly illuminated arena. This absence of population imbalance contrasts sharply with the halved illuminated experiments suggesting that the difference of intensity is behind the stable and systematic accumulation of particles at one side of the arena. 

An alternative way of analyzing the temporal series of Figs.~\ref{Fig:setup_sep}(d)–\ref{Fig:setup_sep}(g) is by means of obtaining a characteristic separation time, denoted $\tau$. This is defined as the time at which the population on the lower-intensity side exceeds, and then consistently remains above $0.5$. Note that, as the population imbalance does not occur for all runs, there are a number of events where we only know that $\tau>360$s. For this reason, and provided that the average is not well defined, in Fig.~\ref{Fig:stationary}(c) we report the known values of $\tau$ together with the median of the measurements (blue diamonds). For $P_{\text{low}} = 23$~mW (darkest blue), all experiments exhibited immediate and clear separation. For $P_{\text{low}} = 33$~mW most of the experiments show quick separation, yielding a median below $20$~s. For $P_{\text{low}} = 43$~mW, $\tau$-s are more disparate, with a median around $3$ minutes. Finally, for $P_{\text{low}} = 52$~mW (i.e. the smallest intensity contrast), the median is around 6 min as only half the experiments displayed clear population separation. The remaining ones resembled the behavior in Fig.~\ref{Fig:setup_sep}(g), with populations fluctuating around $0.5$ and no persistent separation during the duration of the experiment.


\section{Alternating light intensity}\label{Sec:alternating}

\begin{figure}[t]
\centering
\includegraphics[width=0.5\textwidth]{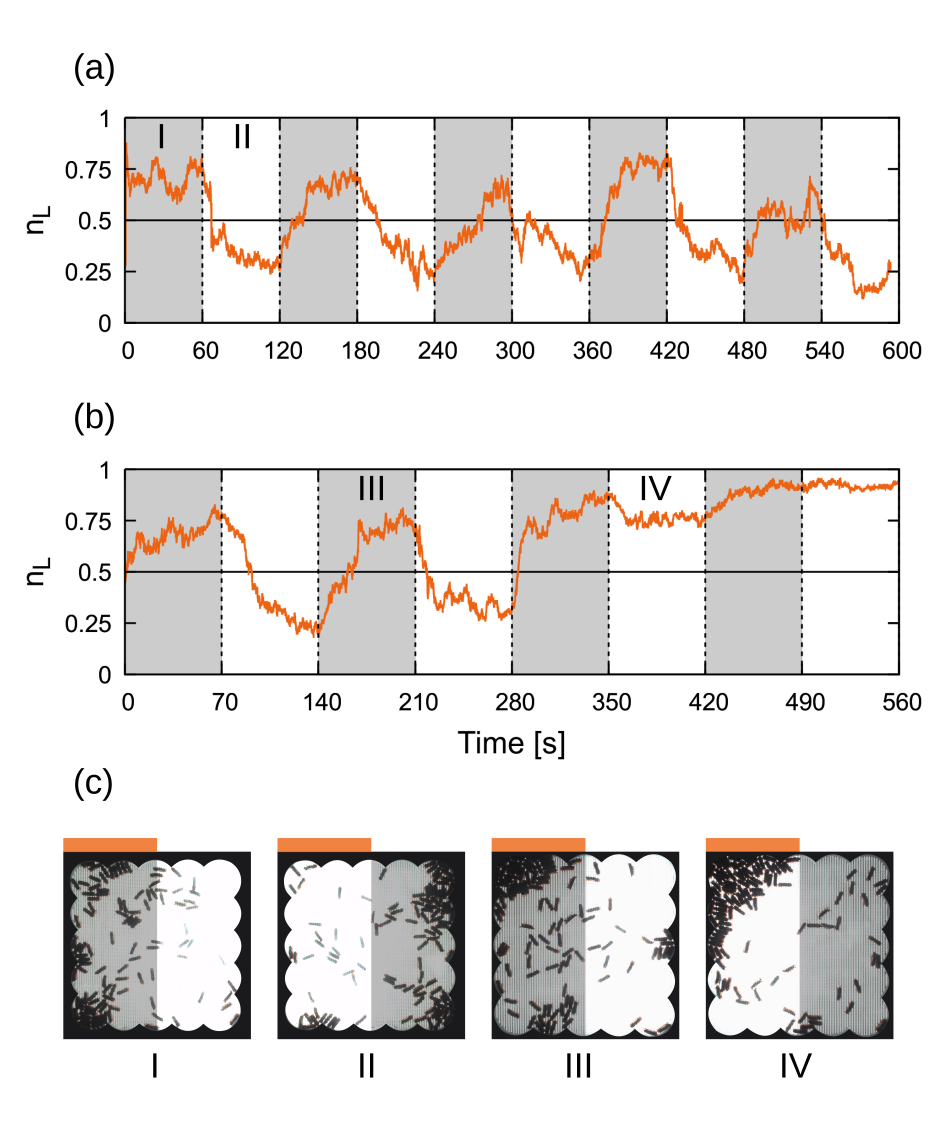}
\caption{\label{Fig:alternating}(a) Relative population ($n_L$) measured at the left-hand side of the system with a switching period of $T{=}60$~s, in a $10$-min-long experiment. The dashed lines mark the different periods of illumination. The gray background represents a period with $P_{\text{low}}$ on the left-hand side. Time intervals I and II correspond to the first two periods whose snapshots are presented in (c). (b) Similar to (a) but for an experiment with $T{=}70$~s switching period. In the first five periods, we have a strong response, particles are accumulating at the low illumination sides (see snapshot III), and in the fifth period, the cluster on the left-hand side stabilizes. After this point, the response stops completely, and the population stays larger on the left-hand side, even when the illumination is high, as shown in snapshot IV. See SM videos 5 and 6~\cite{SM} for the experiments presented in (b) and (c), respectively. (c) Snapshots of the experiments with alternating low and high intensities at the two halves of the system. Note that we apply an additional gray shading on the low-illuminated side of the images to emphasize the difference between sides.}
\end{figure}

Next, we analyze the system response to changes of the activity landscape over time. To this end, we  periodically switch the high ($P_{\text{high}}=72$~mW) and low ($P_{\text{low}}=23$~mW) illumination between the two sides of the arena at every $T=$~\numlist{5;10;20;30;40;50;60;70;80} s. Based on our findings from the stationary case, a preferential accumulation of particles on the darker side can be anticipated; however, it is not clear whether this accumulation will take place in a responsive manner (i.e., following the imposed pattern) for all frequencies explored. Indeed, it can be naively thought that for the highest frequencies the crowd may not have time to migrate from the highest to the lowest intensity side. Differently, for the lowest frequencies, it could also occur that the formation of very large and stable clusters at the dark side prevents their dissolution when the intensity is switched. 

\begin{figure}[t]
\centering
\includegraphics[width=0.5\textwidth]{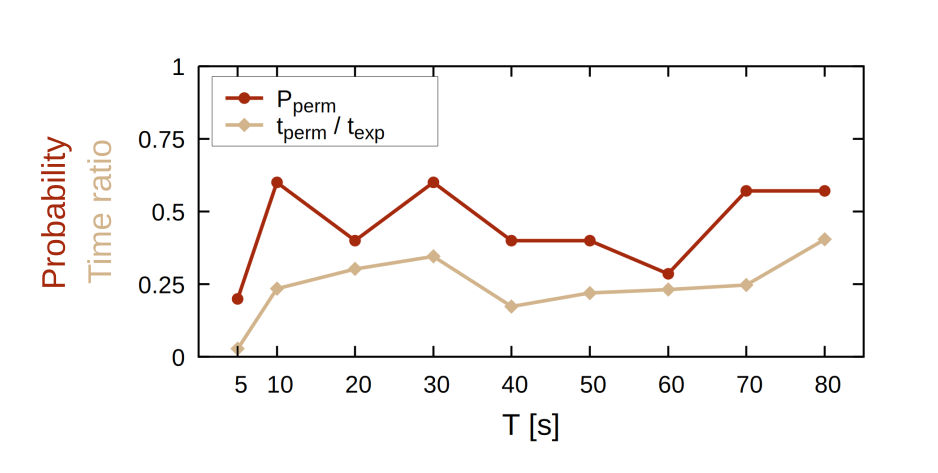}
\caption{\label{Fig:cluster_permanent}Analysis of permanent clusters. Dark red represents the probability that a permanent cluster forms during the $t_{\text{exp}}=10$~min-long experiments as a function of period $T$. Experiments with $T<60$~s were repeated $5$ times, while experiments with $T \geq 60$~s were repeated $6$ times. Light brown represents the proportion of the total experimental time that the system spends in a state of permanent clustering (like snapshot IV in Fig.~\ref{Fig:alternating}).}
\end{figure}

In Fig.~\ref{Fig:alternating} and SM videos 5 and 6~\cite{SM}, we illustrate two of the most typical scenarios observed. In Figs.~\ref{Fig:alternating}(a) and \ref{Fig:alternating}(b), we present the evolution of the number of particles on the left-hand side $N_L$ when $T=60$~s and $T=70$~s, respectively. Note that the gray backgrounds denote periods with lower illumination on the left-hand side. Therefore, in a responsive scenario, these should correlate with higher values of $N_L$, whereas white backgrounds should display lower values of $N_L$. For the case presented in Fig.~\ref{Fig:alternating}(a), the agents accumulate rapidly on the left (darker) side of the arena [snapshot I of Fig.~\ref{Fig:alternating}(c)]. Following the first switch, the particles migrate to the right side, resulting in a decrease in the population on the left [Fig.~\ref{Fig:alternating}(c), snapshot II]. This responsive behavior is maintained throughout the whole experiment, indicating a stable feedback mechanism between illumination and population dynamics. In contrast, in the case of Fig.~\ref{Fig:alternating}(b), different dynamics develop. We observe five responsive periods at the beginning of the experiment [see, for example, snapshot III of Fig.~\ref{Fig:alternating}(c)]. Then, the response stops coinciding with the formation of a large stable cluster which is not destroyed when the light intensity augments in the region of the large cluster [see snapshot IV of Fig.~\ref{Fig:alternating}(c)]. In successive light switches, the accumulation of particles on the left side persists, eventually congregating nearly all of them. From now on, this scenario in which the system falls in a state where a cluster incorporates at least $75\%$ of the particles and does not dissolve throughout the experimental time will be referred to as permanent clustering. There is an alternative scenario where the system does not display permanent clustering but the responsiveness is not totally perfect. This occurs when, during a transient time, $N_L$ does not go above $0.5$ in a darker period or below $0.5$ in a brighter one. After this nonresponsive time during which a permanent cluster is not formed, the system recovers the responsiveness. In what follows we will quantify and analyze these two behaviors and its dependence on the light switching frequency.

\begin{figure}[t]
\centering
\includegraphics[width=0.5\textwidth]{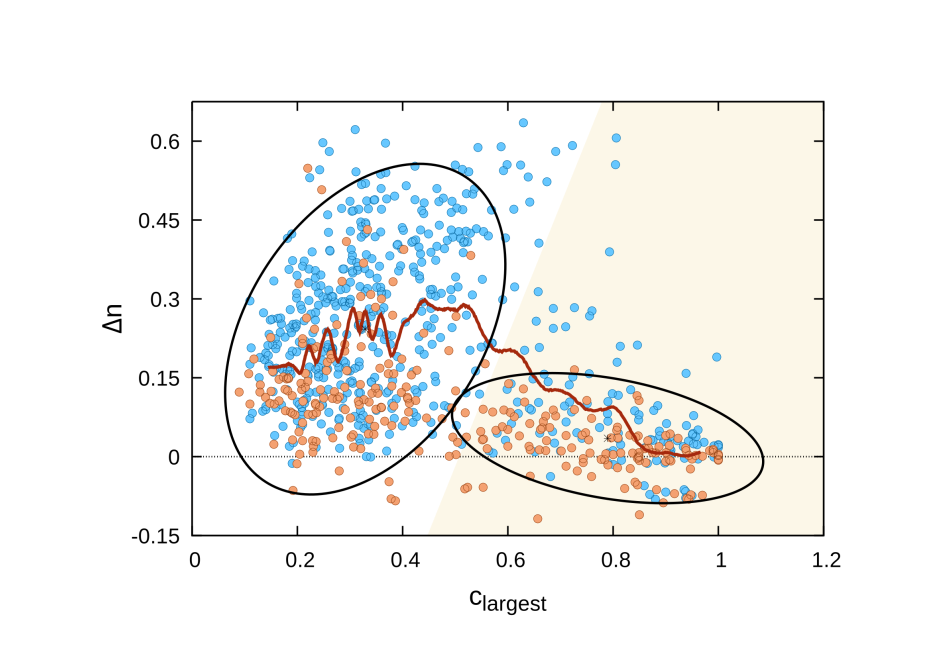}
\caption{\label{Fig:clusters_all}Analysis of the influence of cluster dynamics on system response. $\Delta n$, the increment of the number of agents in the dark side of the arena during a period, is plotted against the relative size of the largest detected cluster, $c_{\text{largest}}$, for experiments with $T > 5$~s. Data points are color-coded: blue indicates that the largest cluster resides in the low-activity region, while orange signifies its presence in the high-activity region of the arena. Two distinct behavioral regimes are identified, highlighted by ellipses and background colors. The right-hand regime is associated with large, stable clusters that exhibit a significantly low dynamic response, whereas the left-hand regime corresponds to smaller and medium-sized clusters that display an enhanced dynamic response.}
\end{figure}

We start by quantifying the prevalence of permanent clustering as a function of the period $T$. To this end, in Fig.~\ref{Fig:cluster_permanent} we show both the probability that an experimental realization ends up in a clustered state ($P_{\text{perm}}$, dark red) and the proportion of experimental time the system remains in the permanent clustering phase (light brown).  Both plots show a similar trend: an initial grow of both magnitudes when going from $T=5$ to $T=10$~s, and then a stabilization for $T>10$~s. This result suggests a rather constant probability of getting a permanent cluster formed, provided that the switching is not too fast.

Next, to better quantify the system responsiveness, we compute $\Delta n$, defined as the difference between the particle populations in the first and last second of a given period at the dark side: $\Delta n = n_{\text{last}}-n_{\text{first}}$. Note that, in principle, $\Delta n$ should be positive if the particles get attracted to the dark side. Then, provided that the presence of large clusters may significantly influence the system response, in Fig.~\ref{Fig:clusters_all} we represent $\Delta n$ as a function of the size of the largest cluster $c_{\text{largest}}$, for experiments with $T > 5$~s. Points are colored blue if the largest cluster resides in the low-activity region, and orange if it is located in the high-activity area. The segregation of the data reveals two distinct regimes, as indicated by the black ellipses, identified by the Gaussian mixture model. A similar separation is identified by k-means clustering, illustrated by the background shading. 

The regime on the left-hand side corresponds to a scenario in which the largest cluster in the arena $c_{\text{largest}}$ agglutinates less than half the population. This regime exhibits an enhanced dynamic response, with $30$\% of the clusters on the high-activity side (orange points) and $70$\% on the low-activity side (blue points). Also, there is a clear positive correlation between $\Delta n$ and $c_{\text{largest}}$ implying that the larger the largest cluster, the more likely that a considerably large group of particles moves from the bright to the dark region. This points toward a global dynamics in which clusters successively form (in the dark region) and destroy (in the bright one). 

\begin{figure}[t]
\centering
\includegraphics[width=0.5\textwidth]{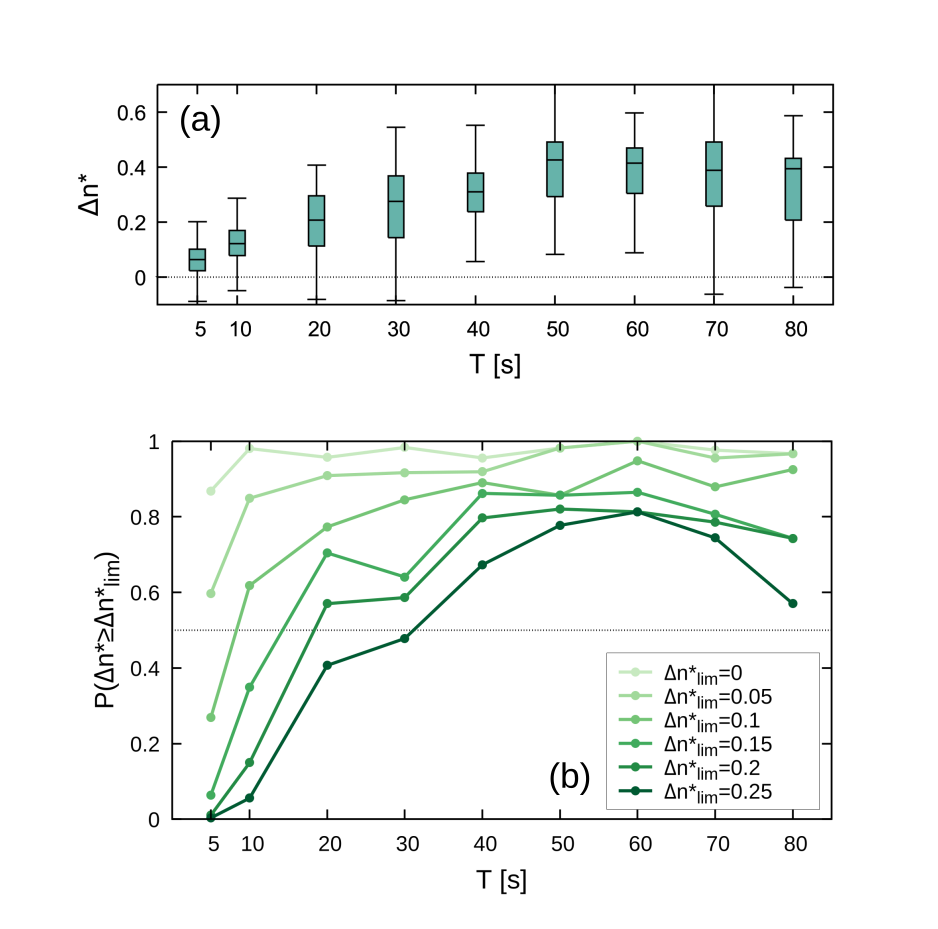}
\caption{\label{Fig:clusters}(a) Box plots of the values of $\Delta n^*$ for different switching periods $T$. (b) Probability of $\Delta n^* \ge \Delta n^*_{\text{lim}}$ as a function of the switching period. Different curves correspond to different values of $\Delta n^*_{\text{lim}}$ as indicated in the legend. The dashed black line corresponds to $P(\Delta n^* \ge 0)=0.5$, representing the expected behavior of a system conformed by randomly moving, noninteracting particles.}
\end{figure}

On the contrary, the regime on the right-hand side of Fig.~\ref{Fig:clusters_all} is mostly characterized by the existence of large clusters (those accounting for more than $60$\% of the total population). This regime is associated with permanent clustering and/or lack of responsiveness, as evidenced by the same number of blue and orange points (thus the same probability of having large clusters at the high and low activity regions). In this regime, $\Delta n$ seems to negatively correlate with the size of the largest cluster $c_{\text{largest}}$, a feature that we attribute to the fact that the larger $c_{\text{largest}}$, the smaller the number of particles moving freely in the arena, so the lower the probability that the amount of particles in the dark side of the arena grows. 

In summary, from Fig.~\ref{Fig:clusters_all} we learn that the presence of large, stable clusters correlates with reduced dynamic response; once established, large clusters maintain their stability independently of their location within the arena (bright or dark regions). In contrast, if the clusters formed at the low-intensity region are small, they tend to dissolve when the intensity increases. In this scenario, larger clusters lead to an increase of the response (measured in terms of $\Delta n$) as indicated by the red line representing the smoothed average of the $\Delta n$ values. The transition among the two regimes (small to medium size dissolving clusters and large permanent clusters) leads to a maximum of $\Delta n$ for $c_{\text{largest}}\approx0.5$; above this value, the risk to get a very large (stable) cluster becomes increasingly important and reduces the system responsiveness.

As the relationship between permanent clusters and response has been established, in the following we restrict the analysis to intervals where permanent clustering has not developed. In Fig.~\ref{Fig:clusters}(a) we represent the system response for these intervals (named $\Delta n^*$ to distinguish from $\Delta n$) as a function of the switching period ($T$). As $T$ grows, both the mean and the span of $\Delta n^*$ grow. Then, for $T\approx50$~s the value of $\Delta n^*$ seems to saturate. Moreover, the box plots of Fig.~\ref{Fig:clusters}(a) reveal an important dispersion of the data, including instances where $\Delta n^*$ is negative; i.e., particles  departing from the low-activity region rather than accumulating there. 

Aiming at a better characterization of the way in which switching frequency affects system response, in Fig.~\ref{Fig:clusters}(b) we represent the probability that within a period, the number of particles accumulated at the dark region exceeds a given threshold  $\Delta n^*_{\text{lim}}$. $\Delta n^*_{\text{lim}}=0$ indicates the simple scenario in which the population of the dark side is larger at the end of the period than at the beginning. For this particular case, the probability of response is almost one for all switching periods, with the only exception of $T=5$ s, for which the probability reduces to $0.85$. This reduction for small values of $T$ becomes more evident when $\Delta n^*_{\text{lim}}$ increases, indicating that very fast switching frequency does not allow the accumulation of particles in the dark side of the arena. As anticipated before, it seems that the curves saturate for $T\approx50$ seconds suggesting that this time should not be overcame when looking for an optimal system response.

All in all, the experimental results show that the responsiveness of the system to the external excitation landscape is conditioned by the presence of clusters. When clusters are small, the population distribution follows the external pattern; i.e., population is larger at the dark region. Indeed, the system responsiveness positively correlates with the cluster size evidencing that clusters continuously form and destroy at each light switching. When clusters are exceedingly large (above half the population), they become stabilized and cannot be destroyed by the increase of energy during the high intensity interval. Then they absorb the majority of the agents in the arena, further increasing the stability and becoming permanent. Clustering is also behind the observed dependence on the switching period. When the period is short, the system is less responsive as clusters do not have time to form and destroy. Increasing the period improves the responsiveness, but this behavior stops when $T\approx50$~s. Above this value, the responsiveness becomes independent of the period and even reduces a little bit; probably due to the formation of very stable, permanent clusters.

As presented in the snapshots of the system [e.g., Fig.~\ref{Fig:alternating}(c)], cluster formation is clearly favored along the system perimeters. This is an expected phenomenon in a finite system where boundaries serve as nucleation points, an effect that is only partially mitigated by the custom arc-shaped (flowerlike) boundary geometry originally designed to reduce this effect~\cite{levay2025cluster}. This boundary-induced stabilization is particularly relevant for the macroscopic agents used here. Recent works have demonstrated that in such systems, the combination of inertial effects and self-alignment can actually suppress the formation of dense clusters in the bulk, potentially promoting flocking or homogeneous phases instead~\cite{caprini2024dynamical, musacchio2025self}. In this context, the boundaries provide the necessary geometric constraints to facilitate the clustering observed under the alternating excitation patterns, counterbalancing the suppressive effects of inertia and self-alignment observed in larger, unbounded systems.

\section{Model}\label{Sec:model}

Aiming for a better understanding of the experimentally observed coupling between clustering and the system’s response to external excitation patterns, we expand a kinetic model that has been recently introduced to explain clustering dynamics under uniform illumination~\cite{levay2025cluster}. In short, the model reproduced the time evolution of the cluster size by considering two processes: adsorption, which accounts for the arrival of new particles to the cluster, and desorption, which accounts for the leaving of particles from the cluster. Under homogeneous illumination, it was found that the stability of clusters was entirely determined by a single kinetic parameter, depending on the density and activity of the agents. The model also predicted a critical cluster size below which the development of stable clusters is not feasible, a feature that was experimentally confirmed posteriorly.

Now, with the aim of reproducing the alternating activity pattern observed in our experiments, it is necessary to account for the interchange of agents traveling between the two reservoirs (left and right side of the arena). In our closed system, the total number of agents $N_T$ is constant, and it includes agents on the left- $N_L(t)$ and right-hand $N_R(t)$ sides, namely, $N_T = N_L + N_R$, and as a result:
\begin{equation}
    \frac{dN_L}{dt} = -\frac{dN_R}{dt}.
\end{equation} 

\noindent Then, within each reservoir, agents can be in two different states: motile free particles ($N_L^f, N_R^f$) or immobile clustered particles ($N_L^c, N_R^c$). Consequently, $N_L = N_L^f+N_L^c$ and $N_R = N_R^f + N_R^c$. The model has two main components: a set of equations describing the interchange between the left- and right-hand sides of the arena, thus determining $N_L$ and $N_R$, and a set of equations accounting for the cluster dynamics, determining $N_L^c$ and $N_R^c$.

To describe the time evolution of the population on each side, we assume that only free agents can move between reservoirs. Thus, the rates of inflow and outflow determine their number. As a first linear approximation, the population change of the left-hand side reads
\begin{equation}
    \frac{dN_L}{dt} = -\alpha_1 N_L^f + \alpha_2 N_R^f,
    \label{eq:transport}
\end{equation}
where the first term represents the decrease in $N_L$ due to free particles leaving the reservoir with rate $\alpha_1>0$, while the second term represents the increase in $N_L$  due to free particles entering the reservoir with rate $\alpha_2>0$. The coefficients, $\alpha_i$, will be referred to as transport coefficients.

For simplicity, we assume that only one cluster develops at a time on each side, and its dynamics is described by the model introduced in Ref.~\cite{levay2025cluster}. Thus, the variation of the cluster size on each side can be written as
\begin{align}
    \frac{dN_L^c}{dt} &= \underbrace{\frac{v_1}{A_{T}}\sqrt{\frac{A_{p}}{\Phi_c}}}_{\beta_1} N_L^f \sqrt{N_L^c} - \underbrace{\frac{D_1\Phi_c}{A_p}}_{\gamma_1} \label{eq:cluster_l}, \\
   \frac{dN_R^c}{dt} &= \underbrace{\frac{v_2}{A_{T}}\sqrt{\frac{A_{p}}{\Phi_c}}}_{\beta_2} N_R^f \sqrt{N_R^c} - \underbrace{\frac{D_2\Phi_c}{A_p},}_{\gamma_2} \label{eq:cluster_r}
\end{align}
with
\begin{align}
    \beta_i = \frac{v_i}{A_{T}}\sqrt{\frac{A_{p}}{\Phi_c}} 
    \text{ and } 
    \gamma_i =  \frac{D_i\Phi_c}{A_p}, \label{eq:betagamma}
\end{align}
where $i \in \{1, 2\}$, and subscripts $1$ and $2$ refer to the low and high intensity sides, respectively. As in \cite{levay2025cluster}, at a given reservoir, $v_i$ is the velocity of the free particles, and $D_i$ is the coefficient of diffusion of particles located at the surface of a cluster. Specific values of $v_i$ and $D_i$ depend on the agents' activity, which is externally controlled by the light intensity at each side. All the other parameters are fixed and related to the experimental conditions. For instance, $A_T$ is the area of the arena available for the agents, $A_p$ is the area of a single particle, and $\Phi_c$ is the packing fraction within a cluster, close to one and constant for simplicity. Therefore, the kinetic parameters $\beta_i$ and $\gamma_i$ fully govern the adsorption and desorption processes, respectively. For the detailed description of the interplay between adsorption and desorption, we refer the reader to Ref.~\cite{levay2025cluster}.

\begin{figure}[t]
\centering
\includegraphics[width=0.5\textwidth]{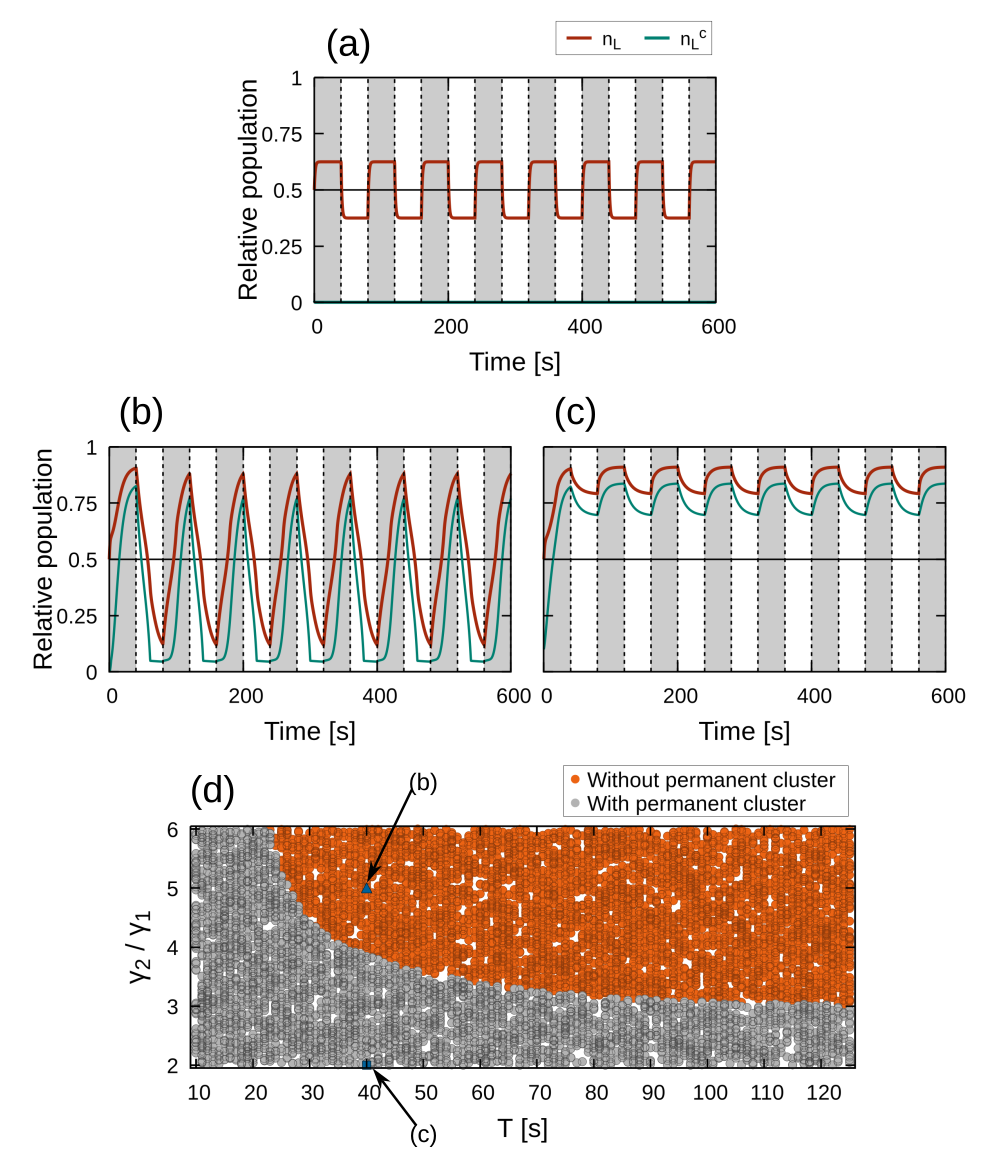}
\caption{\label{Fig:model}Results of the model describing the dynamics of the system with alternating activities. (a)-(c) Temporal evolution of the relative population (red) and cluster size (green) on the left-hand side of the arena when $T=40$~s. Panel (a) is an example in which clustering is nor present ($\beta_i, \gamma_i = 0$) and interchange occurs only due to the difference among $\alpha_1=0.3$ and $\alpha_2=0.5$. (b), (c) Temporal evolution when clustering is present. We use the same transport coefficients $\alpha_i$ as in (a) and we set $\beta_1=0.03$, $\beta_2=0.05$, and $\gamma_1=2$. The only difference among the two plots is the value of the dissolution rate on the high-intensity side, which is $\gamma_2=10$ for (b) and $\gamma_2=4$ for (c). (d) Response of the system for different $\gamma_i$ ratios and switching periods ($T$) for $\alpha_1=0.3$, $\alpha_2=0.5$, $\beta_1=0.03$, and $\beta_2=0.05$. The blue rectangle and triangle indicated by the black arrows represent the case of panel (b) ($T=40$~s, $\gamma_2/\gamma_1=5$) and (c) ($T=40$~s, $\gamma_2/\gamma_1=2$), respectively.}
\end{figure}

To reproduce the experimental periodic switching of the light intensity and the corresponding alternation of particle activity, we periodically switch $\alpha$ (transport coefficient), as well as $\beta$ and $\gamma$ (adsorption and desorption coefficients) between the two sides of the system:
\begin{equation}
    (\alpha_1, \beta_1, \gamma_1) \Leftrightarrow (\alpha_2, \beta_2, \gamma_2) \quad \text{at every } T.
\end{equation}
By convention, we set $\alpha_1<\alpha_2$, $\beta_1<\beta_2$, and $\gamma_1<\gamma_2$, meaning that coefficients with index $1$ represent the low activity, and $2$ the high activity regions. Also, we always start the simulation with the lowest activity at the left-hand side of the arena. 

Based on experimental observations, we constrain the transport and adsorption coefficients by assuming proportionality to the measured particle velocities: $v_1/v_2 = \alpha_1/\alpha_2 = \beta_1/\beta_2$. Also, as it was experimentally observed that $v_1/v_2\approx (8 \text{ cm/s}) / (14 \text{ cm/s})$, we will use this proportion for both, $\alpha_1/\alpha_2$ and $\beta_1/\beta_2$, and will explore the effect of varying the $\gamma_1/\gamma_2$ desorption ratio in the emerging dynamics.

With these premises, we analyze the dynamics of the system over successive switching periods by solving Eqs.\eqref{eq:transport}–\eqref{eq:cluster_r} numerically for different scenarios. First, as a reference, we present the temporal evolution of the population at the left-hand side of the arena for a situation with no clustering; i.e., $\beta_1=\beta_2=\gamma_1=\gamma_2=0$ [Fig.~\ref{Fig:model}(a)]. Clearly, there is a net increase-decrease of the number of particles resulting from a net flux of particles between the two sides of the arena following the imposed period $T=40$~s. This is an obvious consequence of the difference in $\alpha$ coefficients (in this case, $\alpha_1=0.3$ and $\alpha_2=0.5$, but this is robust for any combination of values).

Moving forward to scenarios in which clustering is present, in Figs.~\ref{Fig:model}(b) and \ref{Fig:model}(c) we display two different examples. We set $\alpha_1=0.3$ and $\alpha_2=0.5$ as before, but now we use $\beta_1=0.03$, $\beta_2=0.05$, and $\gamma_1=2$ and vary $\gamma_2$ (the desorption rate at the bright side of the arena). In Fig.~\ref{Fig:model}(b) we show the dynamics for $\gamma_2=10$ (thus, a desorption rate on the high-activity side $5$ times larger than the low-activity one), evidencing a responsive behavior, with a rapid population and cluster size change following every switch. By the end of each low-activity period on the left-hand side (gray regions), around $80$\% of the particles accumulate there, forming a cluster that incorporates most of them. By the end of each high-activity period, however, population decreases drastically, and the cluster size is reduced nearly to zero. This solution of the extended kinetic model resembles the responsive behavior found in experiments, presented in Fig.~\ref{Fig:alternating}(a).

On the contrary, when the desorption rate on the high-activity side is decreased to $\gamma_2=4$, $\gamma_2>\gamma_1$ still holds but a significant fraction of particles become trapped in stable clusters, as shown in Fig.~\ref{Fig:model}(c). The trends of $N_L$ and $N_L^c$ indicate the accumulation of agents on the initially low-activity (left) side. Indeed, accumulation happens already during the first period. Then, although the population and cluster size decreases a little during high-activity periods, the majority of the agents remain on the left-hand side, in the clustered phase. This is analogous to the case of permanent clustering observed in our experiments [Fig.~\ref{Fig:alternating}(b)], where subsequent intensity switches do not significantly affect the situation, and most agents remain trapped in an immobile, permanent cluster.

Next, aiming a most exhaustive analysis of the role of both the ratio among the desorption coefficients $\gamma_2/\gamma_1$ and the switching period $T$, in Fig.~\ref{Fig:model}(d) we present the response of the system as a function of these two parameters. To characterize the response, we first neglect the transient (first $30$\% of the periods). After this time, if the relative population at one side is permanently above $0.5$, we say that the system is in a state of permanent clustering (gray points). Otherwise, if the relative population of both sides alternate above and below $0.5$ following the imposed switching period, we say that the system is responding (orange points). Remarkably, for small values of the switching period $T$, the model predicts a strong dependence of the responsiveness on this parameter. This correlates with the fact that, for small values of $T$, the system becomes non-responsive as clusters may develop but do not have time to destroy. For larger values of $T$, the responsiveness becomes less dependent on this value. In this region, the system becomes responsive as soon as the ratio of the desorption coefficients is above a certain value. This result, compatible with what is shown in Figs.~\ref{Fig:model}(b) and \ref{Fig:model}(c) highlights, once more, the importance of clustering in the collective dynamics of the system.

Importantly, qualitatively similar behavior to the one shown in Fig.~\ref{Fig:model} is obtained for other parameter values. Nevertheless, a further, more detailed and extended study of all possible parameter combinations is pertinent and will be presented elsewhere.   

\section{Conclusion and perspectives}
In this work, we have presented experimental results of the response of a system of macroscopic photoactive particles under several spatiotemporal illuminating fields. In particular, we implement two types of experiments, termed stationary halved illumination and alternating light intensity. In the stationary case, we impose high illumination on the right-hand side and low illumination on the left-hand side. When the contrast between the two halves of the arena is high enough, we observe that the agents tend to migrate toward the darker side. Eventually, the system seems to reach a stationary state with small fluctuations in the relative population on each side. On the contrary, when the difference in illumination between the two sides is small, no sustained population separation is observed.

The analysis of the system response to the alternating light intensity pattern evidences that, generically, the migration patterns can be faithfully controlled; i.e., the concentration of particles follows the externally imposed spatiotemporal pattern moving from the high activity to the low activity region. The system responds to the external spatiotemporal excitation as soon as the switching period is sufficiently long, and no large clustering develops within the arena. Indeed, we discover that the system response is strongly coupled to the cluster development, which strengthens the connection between light intensity and population if clusters are not too large. In this scenario, clusters formed at the low-intensity region dissolve when the intensity increases; hence allowing an effective transport to the other (now darker) side of the arena. On the contrary, the presence of large, stable clusters hinders the dynamic response of the system. Once established, large clusters maintain their stability regardless of whether they lie in the bright or dark region of the arena. 

In summary, our experimental analysis allows for the detection of a transition between two types of behavior, depending on the presence of small dissolving clusters or large permanent clusters. To explain these results, we extended a previous model that describes clustering behavior by a competition between adsorption and desorption processes, and we added a different term that accounts for the transport among the two sides of the arena. Despite its simplicity, the model qualitatively reproduces two of the main experimental features: (i) clustering enhances the strength of the system response [compare Fig.~\ref{Fig:model}(a) with Fig.~\ref{Fig:model}(b)], and (ii) the responsiveness depends on the switching period for small values of this variable and becomes more or less independent for sufficiently large values of it [see Fig.~\ref{Fig:model}(d)]. 

More generally, this work improves our understanding of the dynamics of active matter in landscapes with spatially varying activity, opening possibilities for locally tuning particle density and controlling the transport of active agents. In this context, it is clear that further research is necessary to analyze the system's response in more complex geometrical scenarios (for example, including obstacles) or using external excitations with more sophisticated spatiotemporal dependence (for example, traveling waves or smoother excitation gradients). 

\section*{Acknowledgements}
We especially acknowledge L. Fernando Urrea for technical help. This project has received funding from the European Union’s Horizon 2020 research and innovation program under the Marie Skłodowska-Curie Grant Agreement No. 101067363 named PhotoActive and the Spanish Government through grant No. PID2023-146422NB-I00 supported by MICIU/AEI/10.13039/501100011033. A.K. acknowledges the Asociación de Amigos, Universidad de Navarra, for his grant. AI tools were utilized to improve the style and grammar of the text. The authors are fully responsible and accountable for the content of their article.

\section*{Data availability}
The data that support the findings of this article are openly available~\cite{data}.


%

\clearpage


\onecolumngrid
\begin{center}
  \textbf{\large Supplemental Material for ``Collective dynamics of macroscopic photoactive matter under alternating excitation patterns"}\\[.2cm]
  Sára Lévay,$^{*}$ Axel Katona, Raúl Cruz Hidalgo, Iker Zuriguel\\[.1cm]
  ${}^*$Contact author: slevay@unav.es\\
\end{center}

\setcounter{equation}{0}
\setcounter{figure}{0}
\setcounter{table}{0}
\setcounter{section}{0}
\setcounter{page}{1}
\renewcommand{\thefigure}{S\arabic{figure}}
\renewcommand{\theequation}{S\arabic{equation}}
\renewcommand{\bibnumfmt}[1]{[S#1]}
\renewcommand{\citenumfont}[1]{S#1}
\renewcommand{\thesection}{S~\Roman{section}}

\section{Videos}

\href{https://youtube.com/playlist?list=PLWYWlSkkYOSPoRbTbdxwpFh7cqmIypeRp&si=Ksb2J1QsiBBEcvbl}{SM videos}
\bigskip

\textbf{Video 1} Time evolution of agent populations on both sides of the arena, with $N_T=100$ particles, with stable, halved illumination. The activity on the right-hand side is the highest ($P_{\text{high}}=72$~mW, red), while the activity on the left-hand side is the lowest ($P_{\text{low}}=23$~mW, gray), which still allows the movement of the particles. Starting from a uniform distribution of the agents in the arena, one can observe an immediate separation of the populations; particles are accumulating in the less-illuminated (left, gray) half.
\bigskip

\textbf{Video 2} Time evolution of agent populations on both sides of the arena, with $N_T=100$ particles, with stable, halved illumination. The activity on the right-hand side is the highest ($P_{\text{high}}=72$~mW, red), while the activity on the left-hand side is $P_{\text{low}}=33$~mW. Starting from a uniform distribution of the agents in the arena, one can observe an almost immediate separation of the populations; particles are accumulating in the less-illuminated (left, gray) half, occasionally exhibiting large fluctuations.
\bigskip

\textbf{Video 3} Time evolution of agent populations on both sides of the arena, with $N_T=100$ particles, with stable, halved illumination. The activity on the right-hand side is the highest ($P_{\text{high}}=72$~mW, red), while the activity on the left-hand side is $P_{\text{low}}=43$~mW. Starting from a uniform distribution of the agents in the arena, one can observe the separation of the populations, with high fluctuations, and occasionally equilibration. After some time, a final separation occurs, and most of the particles accumulate on the left-hand side.
\bigskip

\textbf{Video 4} Time evolution of agent populations on both sides of the arena, with $N_T=100$ particles, with stable, halved illumination. The activity on the right-hand side is the highest ($P_{\text{high}}=72$~mW, red), while the activity on the left-hand side is $P_{\text{low}}=52$~mW. As the activity difference is relatively small between the two sides, no sustained population separation is observed; both sides remain close to n=0.5 throughout the experiment.
\bigskip

\textbf{Video 5} Experiment with $N_T=100$ particles with alternating illumination. Starting from $P_{\text{high}}=72$~mW on the right, and $P_{\text{low}}=23$~mW on the left-hand side, illumination is switched at every $T=60$~s. The orange line represents the relative population of the left-hand side. The system shows a response throughout the $10$-minute-long experiment, with particles accumulating at the temporarily dark side.
\bigskip

\textbf{Video 6} Experiment with $N_T=100$ particles with alternating illumination. Starting from $P_{\text{high}}=72$~mW on the right, and $P_{\text{low}}=23$~mW on the left-hand side, illumination is switched at every $T=70$~s. The orange line represents the relative population of the left-hand side. The system shows a response in the first five periods, with particles accumulating at the temporarily dark side, but in the fifth period, a large cluster forms on the left-hand side and gets stabilized. After this point, the response stops completely, and the population stays larger on the left-hand side, even when the illumination is high.
\bigskip

\section{Alternating illumination experiments with $P_{\text{low}}=33$~mW}

In the alternating illumination experiments presented in the paper, we periodically switched the high ($P_{\text{high}}=72$~mW) and low ($P_{\text{low}}=23$~mW) illumination between the sides of the arena at every $T=$~\numlist{5;10;20;30;40;50;60;70;80}~s time. However, we have done the same experimental protocol for $P_{\text{low}}=33$~mW as well, with all other experimental conditions unchanged. In these cases, the activity difference between the two sides is smaller, but still large enough to experience the same phenomena. Here we present all the relevant results of those experiments in the same structure as it is presented in the main paper.

\begin{figure}[t!]
\includegraphics[width=0.5\textwidth]{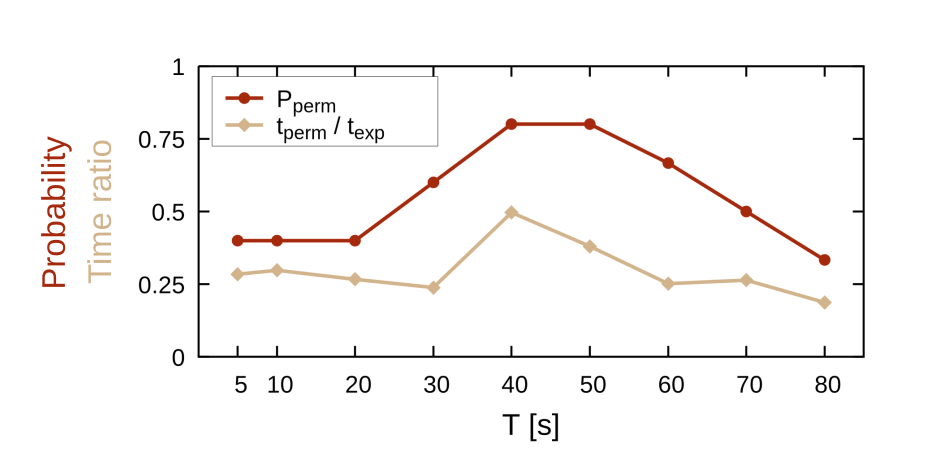}
\caption{\label{Fig:cluster_permanentSM}Analysis of permanent clusters for experiments with $P_{\text{low}}=33$~mW. Dark red represents the probability that a permanent cluster forms during the $t_{\text{exp}}=10$~minute long experiments as a function of period $T$. Experiments with $T<60$~s were repeated $5$ times, while experiments with $T \geq 60$~s were repeated $6$ times. Light brown represents the proportion of the total experimental time that the system spends in a state of permanent clustering. (Cf. Fig.~\ref{Fig:cluster_permanent} of the paper.)}
\end{figure}

Fig.~\ref{Fig:cluster_permanentSM} illustrates the importance of permanent clustering in the dynamics of the system, showing the probability that an experimental realization ends up in a clustered state (dark red), as a function of $T$. The system exhibits a preference for permanent clustering for $T=30-60$~s, but no clear trend is observed for shorter switch periods. When permanent clusters were successfully formed, we quantified the proportion of experimental time the system remained in the permanent clustering phase, represented in light brown.

\begin{figure}[b!]
\includegraphics[width=0.5\textwidth]{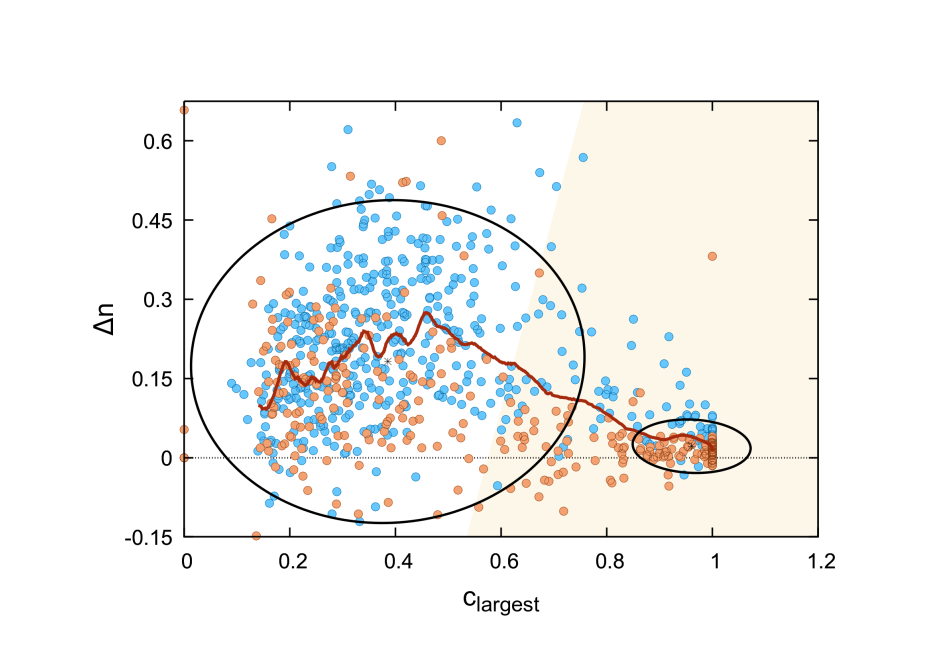}
\caption{\label{Fig:clusters_allSM}Analysis of the influence of cluster dynamics on system response for experiments with $P_{\text{low}}=33$~mW. $\Delta n$, the increment of the number of agents in the dark side of the arena during a period, is plotted against the relative size of the largest detected cluster, $c_{\text{largest}}$, for experiments with $T > 5$~s. Data points are color-coded: blue indicates that the largest cluster resides in the low-activity region, while orange signifies its presence in the high-activity region of the arena. Two distinct behavioral regimes are identified, highlighted by ellipses and background colors. The right-hand regime is associated with large, stable clusters that exhibit a significantly low dynamic response, whereas the left-hand regime corresponds to smaller and medium-sized clusters that display an enhanced dynamic response. (Cf. Fig.~\ref{Fig:clusters_all} of the paper.)}
\end{figure}

Fig.~\ref{Fig:clusters_allSM}, we present the dynamic response as a function of the size of the largest cluster $c_{\text{largest}}$, for experiments with $T > 5$~s. Points are colored blue if the largest cluster resides in the low-activity region, and orange if it is located in the high-activity area. The segregation of the data reveals two distinct regimes, as indicated by the black ellipses, identified by the Gaussian mixture model. A similar separation is identified by k-means clustering, illustrated by the background shading. The regime on the right-hand side is mostly characterized by large, persistent clusters that account for approximately $80$-$90$~\% of the total population. This regime is associated with an exceedingly low dynamic response, evidenced by the same number of blue and orange points in this region, thus the same probability of having these clusters at the high and low activity regions. Conversely, the regime on the left-hand side is dominated by small to medium-sized clusters, which exhibit an enhanced dynamic response, with $30$\% of the clusters on the high-activity side (orange points), and $70$\% on the low-activity side (blue points). As anticipated, the presence of large, stable clusters correlates with reduced $\Delta n$; once established, these clusters maintain stability independent of their location within the arena. In contrast, smaller clusters preferentially form in the low-activity region, leading to the increased dynamic response. As these clusters are inherently less stable, they contribute to a heightened overall response, as indicated by the red line representing the smoothed average of the dynamic response values.

\begin{figure}[t!]
\includegraphics[width=0.5\textwidth]{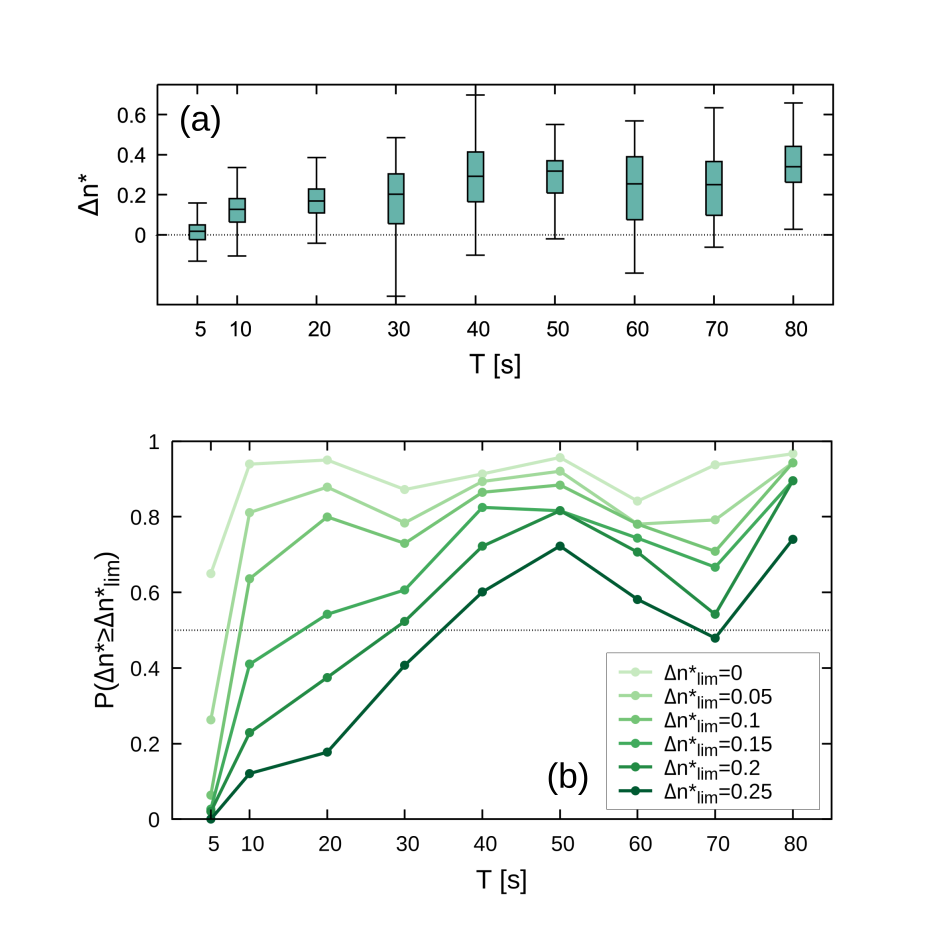}
\caption{\label{Fig:clustersSM}Responsive periods for experiments with $P_{\text{low}}=33$~mW. (a): Boxplots of the values of $\Delta n^*$ for different switching periods $T$. (b): Probability of $\Delta n^* \ge \Delta n^*_{\text{lim}}$ as a function of the switching period. Different curves correspond to different values of $\Delta n^*_{\text{lim}}$. (Cf. Fig.~\ref{Fig:clusters} of the paper.)}
\end{figure}

Fig.~\ref{Fig:clustersSM}(a) illustrates the response of the periods where permanent clustering has not developed. As the switching period grows, both the mean and the span of $\Delta n^*$ grow. Furthermore, we identify instances where $\Delta n^*$ is negative, indicating that in some cases particles are departing from the low-activity region rather than accumulating there.
In Fig.~\ref{Fig:clustersSM}(b), we represent the probability that within a period, the number of particles accumulated at the dark region exceeds a given threshold  $\Delta n^*_{\text{lim}}$. The deeper the color, the stricter the threshold. $\Delta n^*_{\text{lim}}=0$ indicates the simple scenario in which the population of the dark side is larger at the end of the period than at the beginning. In this case, the probability of response exceeds $0.8$ for all switching periods, except $T=5$~s. However, as we increase the threshold, expecting a more substantial difference between the population of the end and beginning of the period (up to $25$\%), we observe that shorter periods tend to exhibit lower responsiveness. In contrast, longer periods demonstrate a higher probability of response. The dashed black line corresponding to $P(\text{response})=0.5$ serves as a benchmark, representing the expected behavior of a system conformed by randomly moving, non-interacting particles.

\end{document}